\documentclass[conference,a4paper]{IEEEtran}

\usepackage[T1]{fontenc}
\usepackage[utf8]{inputenc}

\usepackage[export]{adjustbox}
\usepackage{booktabs}
\usepackage{cite}
\usepackage[dvipsnames]{xcolor}
\usepackage{graphicx}
\usepackage{tabularx}
\usepackage{glossaries}
\usepackage{siunitx}
\usepackage{url}

\makeglossaries

\newacronym{cots}{\mbox{COTS}}{commercial off-the-shelf}
\newacronym{eirp}{\mbox{EIRP}}{effective isotropic radiated power}
\newacronym{gnss}{\mbox{GNSS}}{global navigation satellite system}
\newacronym[plural=\mbox{GPSDOs},firstplural=GPS disciplined oscillators \mbox{(GPSDO)}]{gpsdo}{\mbox{GPSDO}}{GPS disciplined oscillator}
\newacronym{isac}{\mbox{ISAC}}{integrated sensing and communication}
\newacronym{lna}{\mbox{LNA}}{low-noise amplifiers}
\newacronym{los}{\mbox{LoS}}{line-of-sight}
\newacronym{posix}{\mbox{POSIX}}{portable operating system interface}
\newacronym{pps}{\mbox{PPS}}{pulse per second}
\newacronym{rf}{\mbox{RF}}{radio frequency}
\newacronym{rtcm}{\mbox{RTCM}}{radio technical commission for maritime services}
\newacronym{rtk}{\mbox{RTK}}{real-time kinematic}
\newacronym{scpi}{\mbox{SCPI}}{standard commands for programmable instruments }
\newacronym[plural=\mbox{SDRs},firstplural=software defined \mbox{radios (SDR)}]{sdr}{\mbox{SDR}}{software defined radio}
\newacronym{siso}{\mbox{SISO}}{single-input single-output}
\newacronym{snr}{\mbox{SNR}}{signal-to-noise ratio}
\newacronym{ssd}{\mbox{SSD}}{solid-state disk}
\newacronym{ssh}{\mbox{SSH}}{secure shell}
\newacronym{tdoa}{\mbox{TDoA}}{time difference of arrival}
\newacronym{tbv}{\mbox{TBV}}{trust-but-verify}
\newacronym[plural=\mbox{UAVs}]{uav}{\mbox{UAV}}{unmanned aerial vehicle}
\newacronym{ue}{\mbox{UE}}{user equipment}
\newacronym{uhd}{\mbox{UHD}}{USRP hardware driver}
\newacronym{usrp}{\mbox{USRP}}{Universal Software Radio Peripheral}
\newacronym{vpn}{\mbox{VPN}}{virtual private network}

\begin{document}

\title{Measurement Testbed for Radar and Emitter Localization of UAV at 3.75 GHz}

\author{%
	\IEEEauthorblockN{
		Julia Beuster\IEEEauthorrefmark{1}\IEEEauthorrefmark{2},
		Carsten Andrich\IEEEauthorrefmark{1}\IEEEauthorrefmark{2},
		Michael Döbereiner\IEEEauthorrefmark{1},
		Steffen Schieler\IEEEauthorrefmark{2},
		Maximilian Engelhardt\IEEEauthorrefmark{1},\\
		Christian Schneider\IEEEauthorrefmark{2},
		Reiner Thomä\IEEEauthorrefmark{2}
	}
	\IEEEauthorblockA{\IEEEauthorrefmark{1}Fraunhofer Institute for Integrated Circuits IIS, Ilmenau, Germany
	\IEEEauthorblockA{\IEEEauthorrefmark{2}Technische Universität Ilmenau, Institute for Information Technology, Ilmenau, Germany}
	}
}

\maketitle

\begin{abstract}
This paper presents an experimental measurement platform for the research and development of \glspl{uav} localization algorithms using radio emission and reflectivity. 
We propose a cost-effective, flexible testbed made from \gls{cots} devices to allow academic research regarding the upcoming integration of \gls{uav} surveillance in existing mobile radio networks in terms of \gls{isac}.
The system enables nanosecond-level synchronization accuracy and centimeter-level positioning accuracy for multiple distributed sensor nodes and a mobile UAV-mounted node.
Results from a real-world measurement in a 16~km$^2$ urban area demonstrate the system's performance with both emitter localization as well as with the radar setup.

\end{abstract} 


\section{Introduction} 
The steady increase of commercial \gls{uav} applications and the resulting traffic density in the lower airspace are addressed in a defined set of regulations in the so-called U-space, envisaging the integration of \gls{uav} monitoring in mobile radio services (5G and beyond). 
To realize sufficient security against misuse or incorrect use by timely detection of potentially hazardous situations, this cooperative \gls{uav} traffic monitoring system needs to be extended by independent airspace surveillance using radar sensing and radio location, ensuring a \gls{tbv} framework.
The trend in this field is towards making efficient use of the radio resources for \gls{uav} monitoring by extension of existing mobile radio networks in terms of \gls{isac} away from cost-intensive, stand-alone surveillance systems that also lack the flexibility and low-level accessibility needed for researching novel localization algorithms.
Therefore, these commercial solutions are not considered as viable options for an experimental measurement system in this work.
With this paper, we will propose a distributed radio sensor network with nanosecond-level synchronization accuracy, which allows to investigate and develop algorithms for radar and emitter localization of cooperative and non-cooperative \glspl{uav}.
The proposed testbed is modular, tailored for easy deployment and low operational expense and only comprises \gls{cots} devices, to encourage replication in the academic \gls{rf} community.
Recent studies similar to our approach propose \gls{sdr}-based systems for \mbox{radar-only \cite{9087624, 8904605, 8739479}} and emitter-only \mbox{localization \cite{7962829, vtc_droneshield1}}.
Additionally, a comparatively light-weight stand-alone measurement unit tailored for the use at commercial transport \glspl{uav} with a centimeter-level \gls{rtk}-based positioning will be introduced enhancing another recent work that makes use of a relatively heavy-weight \gls{uav}-mounted transmitter in an \gls{sdr}-based channel sounder \cite{9135251}.
Furthermore, we will explain the individual testbed modules and essential integration steps in detail needed to achieve the aforementioned performance and accuracies.
We will validate the testbed's performance exemplarily in a real-world measurement campaign addressing the potential of \gls{isac} for large-scale surveillance in a 16~km$^2$ urban area and small-scale monitoring at a rooftop of a building.
The focus of this publication will be the measurement system itself and not the measurement results.

\section{Measurement System}
\label{section:measurement_system}
\begin{figure}[t]
	\centering
	\includegraphics[width=0.9\linewidth] {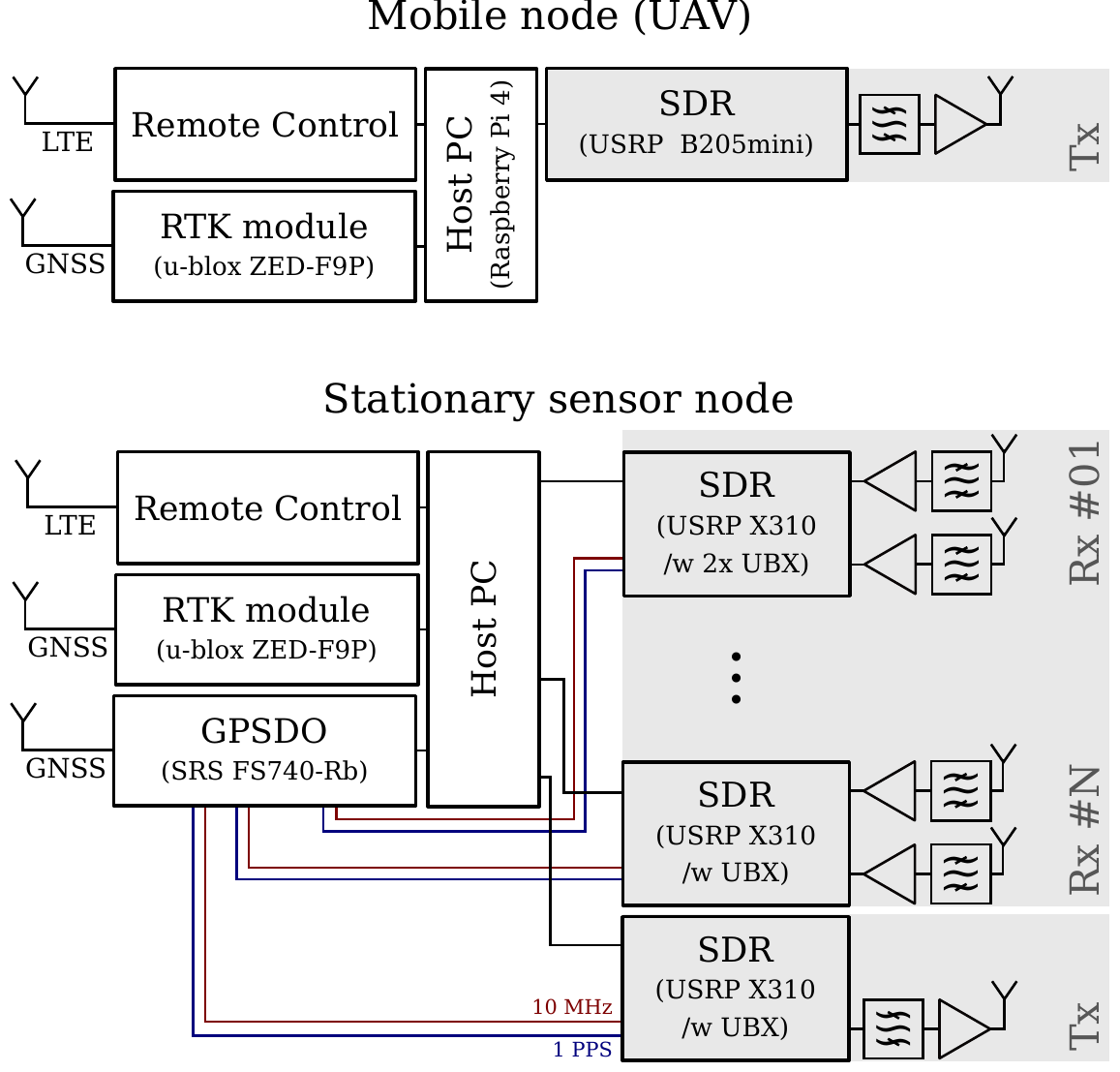}
	\caption{Simplified block diagram of the experimental measurement platform comprised of the light-weight mobile node and stationary sensor nodes. A stationary sensor node consists of one transmitter (radar) and at least one dual-channel receiver (emitter and radar), but their count is only limited by available hardware.
	The equipment part of every stationary sensor node is mounted in 19"~stage racks for easy transportation and deployment to almost arbitrary locations. Each receive and transmit path can be easily activated/deactivated to switch the testbed between radar- and emitter surveillance scenarios.
	}
\label{setup}
\end{figure}

This paper proposes an experimental measurement system for localization of non-cooperative and cooperative \glspl{uav} using radar and radio emissions instead of separate solutions for each use case.
Considering the multitude of localization algorithms, the measurement system prioritizes large bandwidths and highly reliable, continuous data recording with subsequent signal processing for comparing and optimizing localization algorithms without repeating cost-intensive measurement campaigns over real-time localization.
The testbed comprises stationary synchronized distributed sensor nodes tailored for transportability, easy deployment, and low operational expenses, as well as a comparatively small and light-weight stand-alone node with highly accurate ground truth mountable on a common transport \gls{uav}.
The system is modular and only built from \gls{cots} hardware.
Fig.~\ref{setup} shows the key components of the measurement platform.
In the case of emitter localization the sensor nodes act in receive-only mode. Whereas, the localization based on radar requires transmit and receive signal paths.
For both use cases the detection probability and localization accuracy depends on the number of receiving sensor nodes and their geometry.

\subsection{RF Hardware}
The basis of the measurement testbed are \glspl{sdr} from the \gls{usrp} device family which are widely used within the academic \gls{rf} community for a variety of measurement applications due to their flexibility.
This adaptability allows to easily switch the experimental platform between radar- and emitter surveillance scenarios by activating or deactivating individual receive and transmit paths.

\subsubsection{Mobile node}
\begin{figure}[t]
	\centering
	\includegraphics[width=0.45\linewidth] {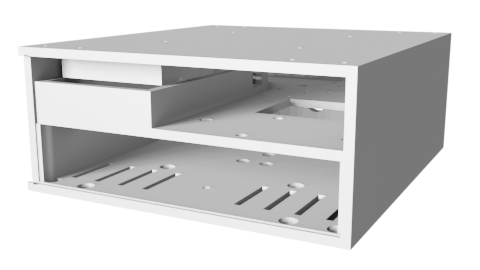}
	\includegraphics[width=0.45\linewidth] {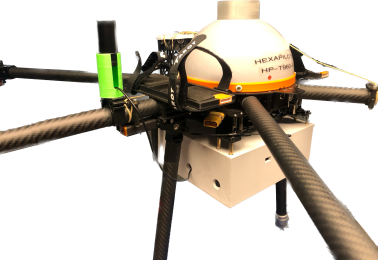}
	\caption{3d printed protective housing to attach mobile node components to a transport \gls{uav}, exemplarily the Tarot 960 Hexacopter.}
\label{uav}
\end{figure}

With regard to the strict weight and space limits when operating \glspl{uav}, the mobile measurement unit is equipped with the \gls{sdr} transceiver platform \gls{usrp} B205mini compromising between a small form factor and sufficient bandwidth of up to 56 MHz.
It is extended by an additional band-pass filter to minimize 
spurious radio emissions, a light-weight power amplifier, and a relatively small omni-directional antenna to achieve up to 23 dBm \gls{eirp}.
A small host PC, namely the Raspberry Pi 4, streams the pre-configured baseband signal via USB 3.0 to the \gls{sdr}.
To use the mobile node as stand-alone unit and avoid repetitive system
warm-up periods in between flights, all components are powered by one self-sufficient \mbox{5V power supply} in form of an \mbox{400 g exchangeable} powerbank instead of using the battery of the \gls{uav} allowing operation times of up to \mbox{5 hours}.
All components are installed in a customized \mbox{3d printed} protective housing with dimensions measuring 200 x 210 x 85 mm resulting in an overall node weight of \mbox{800 g}. The housing can be attached to a transport \gls{uav}, exemplarily the Tarot 960 Hexacopter with an airtime of up to 25 minutes, as shown in Fig.~\ref{uav}.

\subsubsection{Stationary sensor node}
The key component of the distributed sensor nodes is the dual-channel \gls{sdr} transceiver platform \gls{usrp} X310 with UBX frontends.
Each \gls{sdr} is connected via \mbox{10 GBit Ethernet} to a high-performance host server  equipped with arrays of \gls{ssd} for signal storage to stream the baseband signals from/to.

The transmitter employs a band-pass filter to minimize harmonics and spurious emissions, an additional power amplifier and a high gain antenna to achieve \mbox{46 dBm \gls{eirp}}.

All receive paths are equipped with band-pass filters and subsequent \gls{lna} to minimize noise figure and effects from strong interferers.
To reduce attenuation, the cable between antenna and amplifier is as short as possible.
All these factors create the need of multiple \glspl{sdr} to increase the number of sensor positions and therefore the probability of detection and localization accuracy.
To use different antenna types at one receiver position, exemplarily one omni-directional and one directional antenna, the \gls{sdr} is operated as dual-channel receiver.

The hardware of each sensor node can be mounted in 19" racks to support transportability and distributed deployment.

The proposed measurement system currently uses four stationary sensor nodes, each comprising one transmitter, one dual-channel receiver with an omni-directional and a directional antenna, and two single-channel receiver nodes with directional antennas, but their count and setup is only limited by available hardware.

\subsection{Distributed Synchronization}
The distributed, stationary sensor nodes must be accurately synchronized to enable reliable localization results.
Therefore, each node employs a Stanford Research Systems FS740 \gls{gpsdo} with internal Rubidium reference for straightforward synchronization at almost arbitrary locations using an internal Trimble RES SMT 360 \gls{gnss} timing receiver, only requiring a clear sky view.
The difference between the \gls{gnss}' \mbox{1 \gls{pps}} signal and the \gls{gpsdo}'s time base is used as PLL input and also accessible via \gls{scpi} commands.
Within one stationary sensor node all \glspl{sdr} are synchronized via \mbox{10 MHz} and 1 \gls{pps} reference signals from one \mbox{SRS FS740}.

As we've previously elaborated, the SRS FS740s' \mbox{10 MHz} outputs are not phase aligned.
They exhibit a random initial phase offset that changes with each power cycle~\cite[Fig. 2]{vtc_droneshield1}.
The USRPs' sampling clocks, which are derived from the \mbox{10 MHz} signals, inherit that phase offset.
Additionally, different cable lengths, thermal factors, and other analog components will also impact the absolute synchronization.

All time-invariant synchronization errors can be measured and subsequently compensated in post-processing by using a temporary beacon transmitter with unobstructed \gls{los} to all receivers at an exactly determined location measured using an \gls{rtk} \gls{gnss} with cm-level accuracy (see following Sec.~\ref{section:gnss}).
To calibrate and repeatedly verify the synchronization between all receiver nodes, the stationary radar transmitter is intermittently deployed at such locations.

\begin{figure}[t]
	\centering
	\includegraphics[width=1\linewidth]{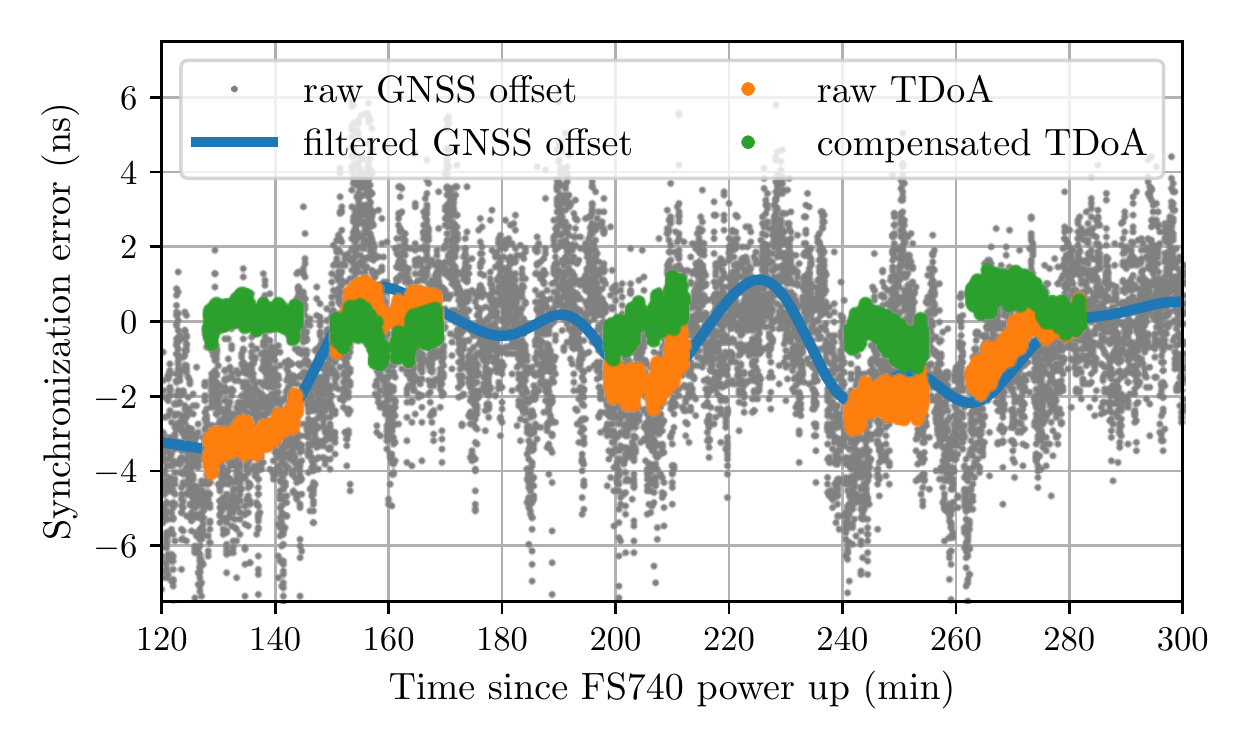}
	\caption{%
		Synchronization error between receivers at location \#01 and \#02 determined via \gls{tdoa} of the beacon transmitter with \gls{los} to both receivers.
		The beacon is used intermittently, so \gls{tdoa} is not available continuously.
		Raw \gls{tdoa} correlates with raw measurements of the \gls{gnss} timing receivers of \mbox{SRS FS740} \glspl{gpsdo} used to synchronize all stationary sensor nodes.
		By post-processing the raw \gls{gnss} time error, the \gls{tdoa} variance drops from $ (\SI{1.27}{\nano\second})^2 $ to $ (\SI{0.482}{\nano\second})^2 $.
	}
\label{fig:sync}
\end{figure}

Fig.~\ref{fig:sync} illustrates the time-variant synchronization error between receivers at location \#01 and \#02 (cf. Fig. \ref{scenario}), which are separated by 1.25~km.
This error was determined via high-resolution parameter estimation of the \gls{los} path propagation delay (see Sec.~\ref{section:radar_hrpe} for details) for each receiver.
The drift caused by the temporary transmitter is eliminated by computing the difference between two receivers, i.e., the \gls{tdoa}.
The variance of the beacon's uncompensated \gls{tdoa} in Fig.~\ref{fig:sync} is $ (\SI{1.27}{\nano\second})^2 $.
Fig.~\ref{fig:sync} illustrates an obvious correlation between the measured \gls{los} \gls{tdoa} error and raw \gls{gnss} time differences between the \mbox{SRS FS740} in use.
To further improve the synchronization accuracy, the \mbox{SRS FS740s'} \gls{gnss} receiver data is post-processed via an empirically derived rectangular low-pass filter, reducing the \gls{los} \gls{tdoa} variance to $ (\SI{0.482}{\nano\second})^2 $.

\subsection{GNSS Ground Truth}
\label{section:gnss}
Highly accurate positioning with cm-level accuracy is required for synchronization, localization, and as ground truth for verifying localization results.
This applies to static locations of stationary antennas as well as the dynamic position of the \gls{uav}.
We rely on low-cost \gls{rtk} dual-band \gls{gnss} receivers, namely u-blox ZED-F9P, in combination with real-time \gls{rtcm} correction data from the Thuringian SAPOS\textsuperscript{\textregistered} system \cite{sapos}.
The solutions enables a theoretic accuracy of \SI{1.8}{\centi\meter} as reported by the \gls{gnss} receivers under ideal conditions. 

Unfortunately, the actual accuracy is difficult to determine without a professional, expensive inertial navigation system.
To estimate the ZED-F9P's positioning performance, we placed two \gls{gnss} antennas \SI{1.65}{\meter} apart on a car roof.
Because the antennas' positions on the roof are fixed, the distance between both antennas measured via \gls{gnss} must be constant and equal to the geometrically measured distance between both antennas' phase centers.
Any deviations are caused by the \gls{gnss} receivers and/or the \gls{gnss} propagation channel, i.e., non-line-of-sight and multipath conditions.
Since the antennas are placed more than 6 wavelengths apart, the effect of multipath propagation will substantially differ between both.

Driving in the town of Ilmenau, in between multi-story buildings guaranteeing strong multipath environments (cf. Fig.~\ref{flighttrajectories}), we measured a 90\% quantile with less than \SI{3}{\centi\meter} and a 99\% quantile with less than \SI{35}{\centi\meter} baseline length error.
Multipath propagation is negligible for the \gls{uav} and our chosen static antenna locations, as all have an unobstructed view of hemisphere.
Therefore, we are highly confident that our absolute \gls{gnss} position accuracy is better than \SI{3}{\centi\meter}.
 
\subsection{Software}
As motivated in Sec.~\ref{section:measurement_system}, the measurement system is tailored to record data for offline processing and algorithm research instead of real-time localization.
We developed a suitable C++ software using the \gls{uhd} library.
This enables coherent sampling using multiple spatially distributed USRPs.
Most importantly, loss of Ethernet frames is transparently accounted by padding the output file with the correct amount of zero-valued samples to avoid zero byte gaps in the sample stream.
Without the applied zero padding, loss of a single Ethernet frame would cause irreparable loss of coherence between files.

To support easy deployment and low operational effort, the software enables remote control and telemetry via straightforward \gls{posix} signal and shared memory interfaces.
Via LTE modems, a \gls{vpn}, and \gls{ssh} remote console access, the exemplary setup used in the real-world measurement campaign (see Sec.~\ref{section:meas_campaign}) consisting of 8 USRPs at 5 different locations was conveniently orchestrated by a single operator from an office workstation.
Measurement data, primarily channel impulse responses at all receivers, was available in real-time to continuously monitor and validate the measurement.

The \gls{usrp} measurement software is available under GPLv3 license from our GitHub account~\cite{github_usrp_rxtx}.

\section{Measurement Campaign}
\label{section:meas_campaign}
\begin{figure}[t]
	\centering
	\includegraphics[width=1\linewidth] {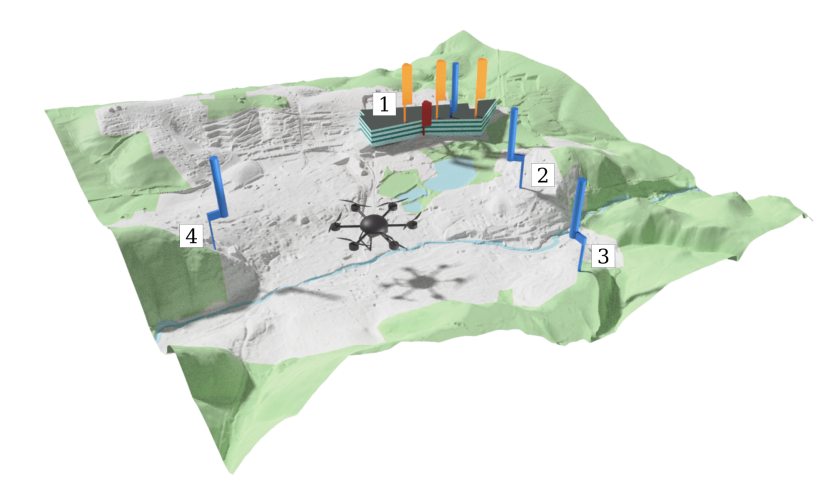}
	\caption{Simplified illustration of the real-world measurement setup to examine radar- and emitter localization of \glspl{uav}. The chosen scenario consists of one mobile measurement unit mounted to an \gls{uav} with reliable ground truth and four synchronized, stationary sensor nodes deployed in a \mbox{4\,km by 4\,km} urban area. Each stationary unit has at least one activated receive path equipped with an omni-directional antenna \mbox{({\color{MidnightBlue} $\bullet$})}. Additionally, the sensor node at \mbox{location \#01} is setup using directional antennas at three receive paths \mbox{({\color{BurntOrange} $\bullet$})} and one transmit path \mbox{({\color{BrickRed} $\bullet$})}. The map data is made available by \cite{geoportal}.}
\label{scenario}
\end{figure}

To demonstrate the level of flexibility offered by the proposed experimental testbed regarding the development of algorithms for radar and emitter localization in terms of \gls{isac}, we conducted an exemplary real-world measurement in an urban scenario with one mobile \gls{uav}-mounted node and four stationary sensor nodes, as depicted in Fig.~\ref{scenario}.
The sensors were distributed all over the city of Ilmenau, Germany, with exposed antenna positions similar to base stations, for instance at roof tops and tripods.
The setup is chosen with at least one receiver (Rx\#01) at each location equipped with an omni-directional antenna, exemplarily for large-range emitter localization using \gls{tdoa}-based multilateration.
Additionally, one sensor node location is chosen to exemplify small-range radar localization.
Therefore, the stationary node at \mbox{location \#01} is setup on the rooftop of a four-story building, as shown in Fig.~\ref{scenario}, with one dual-channel receiver (Rx\#01) using both an omni-directional and a directional antenna, two receivers (Rx\#02, \#03) each equipped with a directional antenna, one transmitter, and additional absorbers to minimize \gls{los} power and therefore, to improve receiver sensitivity and linearity.

\begin{figure}[h]
	\includegraphics[width=0.59\linewidth] {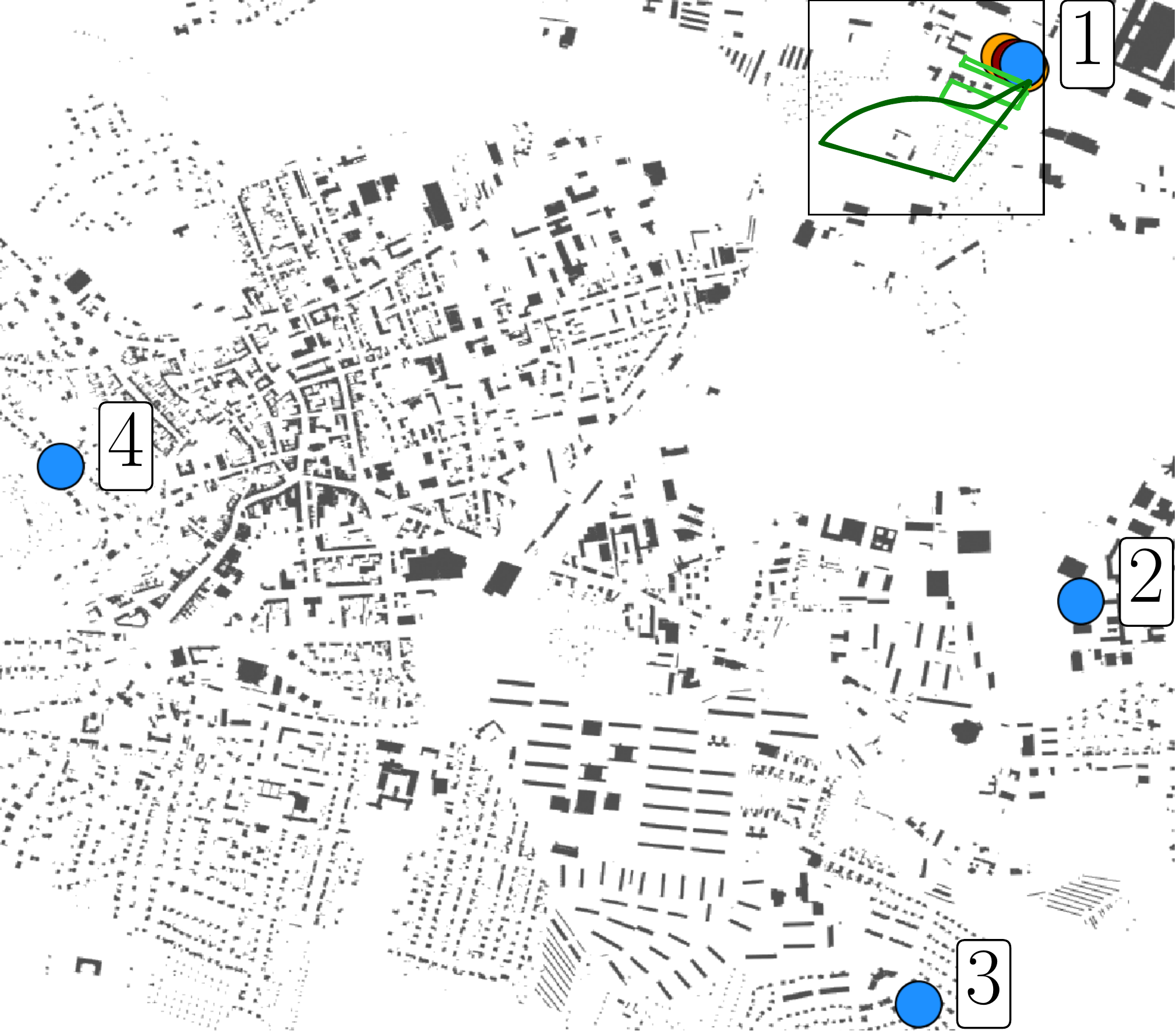}
	\includegraphics[width=0.4\linewidth] {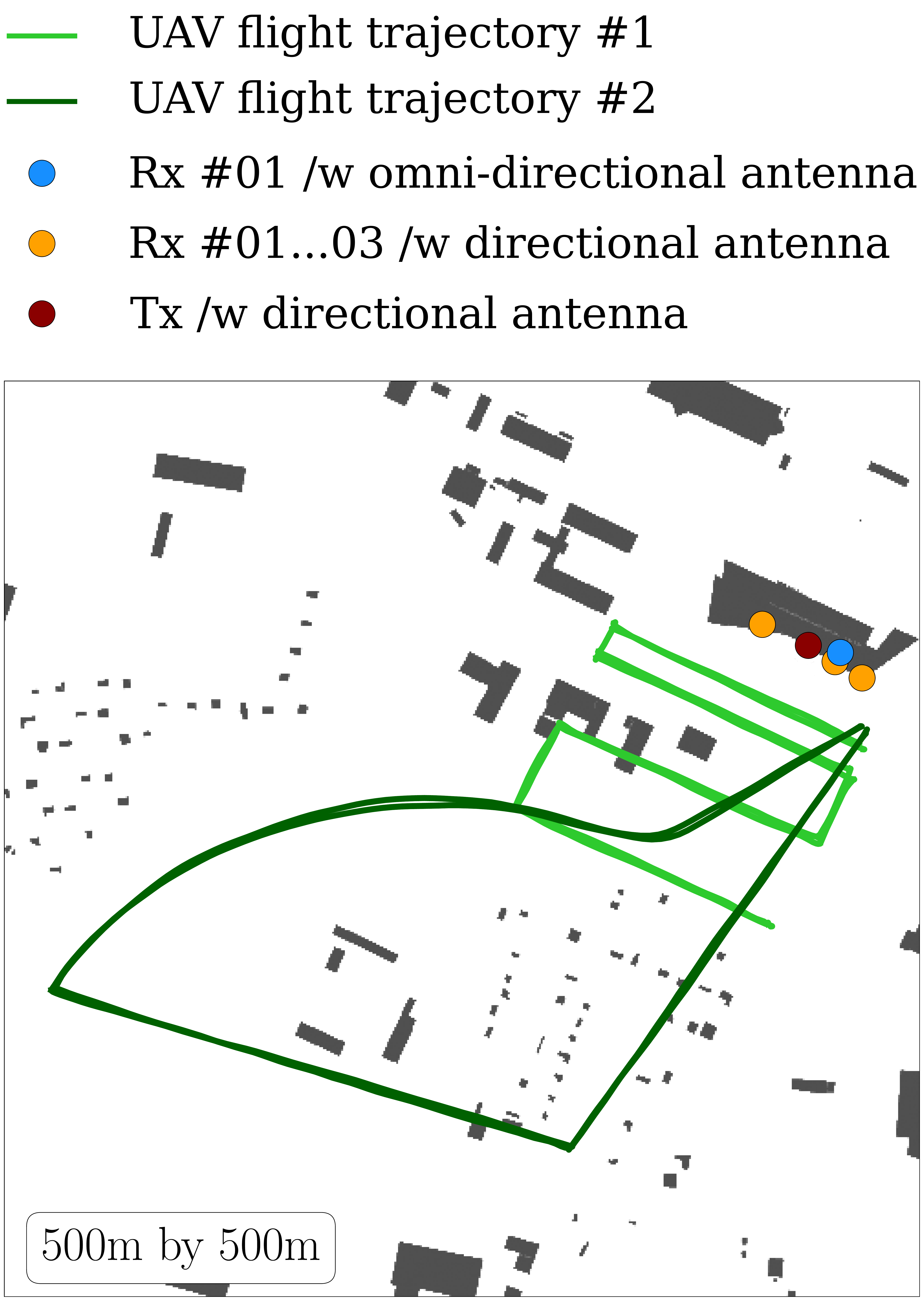}
	\caption{Flight route of the \gls{uav} including a combination of parallel and perpendicular flight in arrival and departure situations. The map data is made available by \cite{geoportal}.}
\label{flighttrajectories}
\end{figure}

The flight trajectories of the \gls{uav} were planned to include a combination of parallel and perpendicular flight in arrival and departure situations, as shown in Fig.~\ref{flighttrajectories}, with varying flight velocities and varying flight altitudes above and below system level to generate a diverse data set.

Due to the upcoming rise of \gls{uav} integration in 5G, the scenario was tailored to provide conditions and signal parameters similar to those present in the localization of \mbox{5G \gls{ue}} in FR1 \cite{3gpp.36.211}, as shown in Table~\ref{tab:specs}.

\begin{table}[h]
	\caption{Summary of measurement setup characteristics.}
	\vspace{-1em}
	\centering
    \begin{tabularx}{0.95\linewidth}{Xl}
	\toprule
	\small Real-world measurement campaign\\
	\quad -- Urban area & $\sim$ 16 km$^2$\\
	\quad -- Mobile node & 1, \gls{uav}-mounted \\
 	\quad -- Stationary sensor nodes & 4\\ 
	\quad -- \gls{rtk} base station & SAPOS\textsuperscript{\textregistered}\\
	\midrule
	\small Radar Surveillance testbed & \\ 
	\quad -- mobile Tx (\gls{uav}) & muted \\
	\quad -- stationary Tx\footnotemark \mbox{({\color{BrickRed} $\bullet$})}& 1 \\
	\quad -- Rx /w directional antenna\footnotemark[1] \mbox{({\color{BurntOrange} $\bullet$})} & 3 (location \#01) \\
	\quad -- Rx /w omni-directional antenna \mbox{({\color{MidnightBlue} $\bullet$})} & -- \\
 	Signal parameters & \\
 	\quad -- OFDM, Newman phase sequence\\
 	\quad -- Center frequency & 3.75 GHz \\
 	\quad -- Bandwidth (used) & 80 MHz \\
 	\quad -- Subcarriers & 1280 \\
	\quad -- Symbol length & 16 \textmu s \\
 	\quad -- Effective isotropic radiated power & 46 dBm\\
 	\midrule
	\small Emitter Surveillance testbed & \\
	\quad -- mobile Tx (\gls{uav}) & 1 \\
	\quad -- stationary Tx \mbox{({\color{BrickRed} $\bullet$})} & muted \\
	\quad -- Rx /w directional antenna\footnotemark[1] \mbox{({\color{BurntOrange} $\bullet$})}& --\\ 
	\quad -- Rx /w omni-directional antenna \mbox{({\color{MidnightBlue} $\bullet$})} & 4 (location \#01...\#04) \\
 	Signal parameters & \\
 	\quad -- OFDM, Newman phase sequence\\
 	\quad -- Center frequency & 3.75 GHz \\
 	\quad -- Bandwidth (used) & 32 MHz \\
 	\quad -- Subcarriers & 512 \\
	\quad -- Symbol length & 16 \textmu s \\
 	\quad -- Effective isotropic radiated power & 23 dBm\\
 	\bottomrule
	\multicolumn{2}{l}{\footnotemark[1] \scriptsize Directional antenna with 40 degree 10dB beamwidth at 3.75 GHz.}\\
 	\end{tabularx}
 	\label{tab:specs}
 \end{table}

\subsection{Radar Surveillance}
\label{section:radar_hrpe}
To testify localization of \glspl{uav} that do not emit any signals, the testbed was operated in radar mode with a muted transmitter at the \gls{uav}.
Since the radar range coverage is lower than for localizing an active radio emitter, only the stationary sensor node at location \#01 is used with one stationary transmitter and three stationary receivers, as shown in Table~\ref{tab:specs}, with an estimated range of up to \mbox{230 m}.

Due to the wide assortment of radar-based localization approaches, the evaluation of particular algorithms based on the measured data is not in the scope of this paper.
The system's capability to detect an \gls{uav} using its reflectivity is demonstrated using a simple \gls{los} situation at a flight altitude of \mbox{30 m above} ground.
To estimate the \gls{uav} position, an averaging of \mbox{20 consecutive} snapshots is performed to increase the \gls{snr}.
Subsequently, a delay line canceler is applied to suppress the influence of static clutter \cite{1187759}.
The delay parameter of the target reflection is obtained using a maximum-likelihood based parameter estimator.
Afterwards, multi-target tracking based on a Kalman filter and Hungarian Matching \cite{6866690} is used to track the delay parameter of the target and discard remaining clutter estimates.
Finally, the estimated target delays of all receiver nodes are fused together with the position information of the transceiver nodes to calculate the target positions using least-squares.

As shown in Fig.~\ref{localizationerror}, the \gls{uav} is detected in \mbox{60 percent} of the flight route, mainly limited by the antenna sector. 
Missing positions are outside of \mbox{10 dB beamwidth} of directional antennas \mbox{(cf. Table~\ref{tab:specs})} which measures up to \mbox{20 dB} power loss of the backscattered signal from the target.

\subsection{Emitter Surveillance}
To demonstrate the localization of \glspl{uav} based on radio emissions, the stationary sensor nodes at location \mbox{\#01...\#04} are operated in receive-only mode. 
Because of the vast variety of available localization techniques using radio emissions this paper focuses only on a simple geometrical approach using \gls{tdoa}-based multilateration with omni-directional antennas \mbox{(cf. Table~\ref{tab:specs})} in a \gls{los} situation between the \gls{uav} transmitter and all four receiver nodes with a flight altitude of \mbox{30 m above} ground.
The examination of individual algorithms based on the measurement data is not in the scope of this paper.
The demonstrated estimation of unknown \gls{uav} emitter positions is based on computing the \gls{tdoa} using pairwise cross-correlation of the received signals and a simple least squares method to solve the hyperbolic equations \cite{shen2008performance}.

\begin{figure}[t]
	\centering
	\includegraphics[width=0.8\linewidth, valign=c] {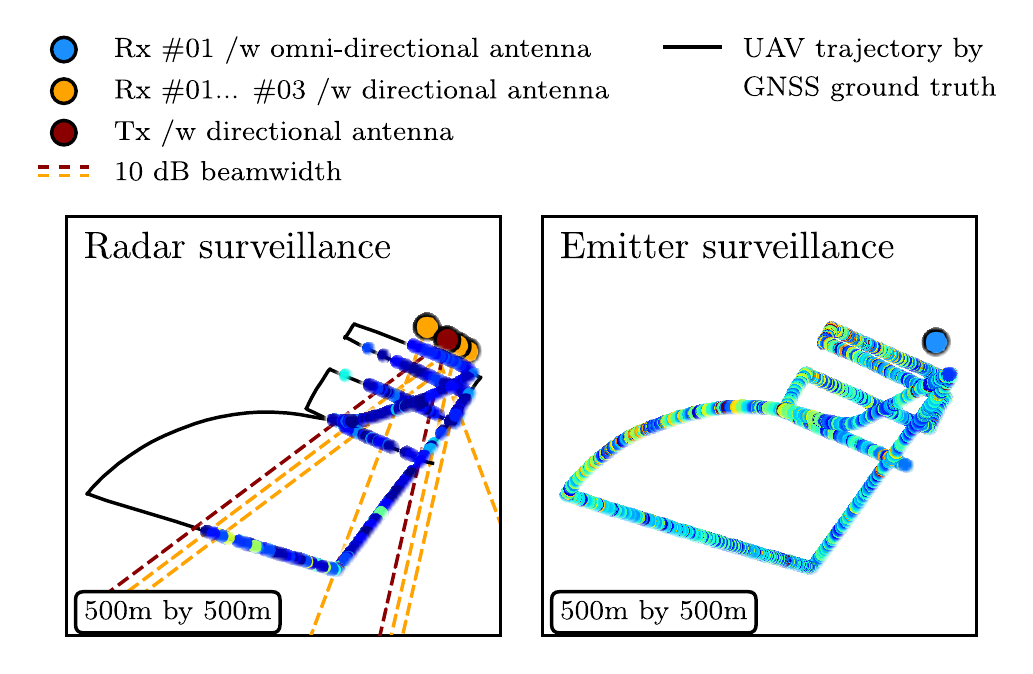}
	\includegraphics[width=0.15\linewidth, valign=c] {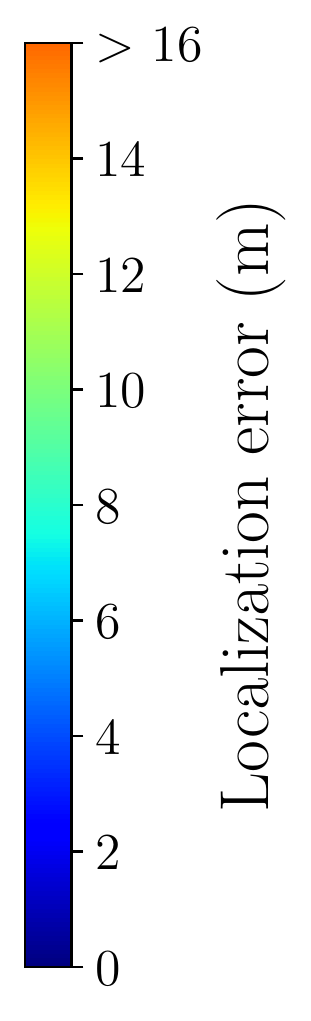}
	\caption{Localization error between actual and estimated \gls{uav} position in \gls{los} situation at a flight altitude of \mbox{30 m above} ground using the experimental testbed in radar-only (left) and emitter-only mode (right).}
\label{localizationerror}
\end{figure}

As shown in Fig.~\ref{localizationerror}, the \gls{uav} is detected along the  flown tracks. 
Due to the geometrical arrangement of the four sensor nodes all over the city and the simplicity of the used localization technique, the overall localization error is slightly larger compared to the small-range radar surveillance.

\section{Conclusion}
In this research we have introduced a distributed measurement testbed enabling the localization of actively emitting \glspl{uav} with a radio emission-based approach as well as the localization of non-emitting \glspl{uav} using their backscattered signals in a radar-based use case.
The measurement system is based exclusively on \gls{cots} components and covers distributed stationary sensor nodes with nanosecond-level synchronization accuracy and a light-weight mobile stand-alone unit, both with an \gls{rtk}-based positioning with centimeter-level accuracy.
By demonstrating the system's performance with an exemplary real-world measurement in a 4km by 4km urban scenario using a commercial transport \gls{uav}, we have addressed implementation details and challenges of assembling a distributed sensor network in order to enable replication and encourage the academic \gls{rf} community to validate future emitter and radar localization algorithms by real-world data in addition to relying on simulations.

\section*{Acknowledgements}
The research has been partially funded by the Federal State of Thuringia, Germany, and the European Social Fund (ESF) under the grant 2018 FGR 0082.
The authors acknowledge the financial support by the Federal Ministry of Education and Research of Germany in the projects "6G-ICAS4Mobility" (grant number: 16KISK241) and "KOMSENS-6G" (grant number: 16KISK125).
We thank Dr. Axel Weckschmied (HEXAPILOTS\textsuperscript{\textregistered}) and Uli Barth (C5UAV GmbH) for supporting the measurement campaign this publication is based on by providing and operating the \gls{uav}.

\bibliographystyle{IEEEtran}
\bibliography{paper}
\end{document}